\documentstyle{l-aa}
\newcommand{\etal}{{\em et al.~}}
\newcommand{\ce}{Cepheid~}
\newcommand{\cs}{Cepheids~}
\newcommand{\kms}{km.s$^{-1}~$}

\begin{document}
\thesaurus{08.06.3,08.05.2,12.04.1,12.07.01}
\title{Metallicity Effects on the Cepheid Extragalactic Distance Scale from
EROS photometry in LMC and SMC} 

\author { 
D.D. Sasselov\inst{1,2}, J.P.Beaulieu\inst{2} C. Renault\inst{3}, P. Grison\inst{2}, 
R.Ferlet\inst{2}, A. Vidal-Madjar\inst{2}, E. Maurice\inst{5}, L. Pr\'evot\inst{5}, 
E. Aubourg$^{3}$, P. Bareyre\inst{3}, S. Brehin\inst{3}, C. Coutures\inst{3}, 
N. Delabrouille\inst{3}, J. de Kat\inst{3}, M. Gros\inst{3}, B. Laurent\inst{3},
 M. Lachi\`eze-Rey\inst{3}, E. Lesquoy\inst{3}, C. Magneville\inst{3}, 
A. Milsztajn\inst{3}, L. Moscoso\inst{3}, F. Queinnec\inst{3}, 
J. Rich\inst{3}, M. Spiro\inst{3}, L. Vigroux\inst{3}, S. Zylberajch\inst{3}, 
R. Ansari\inst{4}, F. Cavalier\inst{4}, M. Moniez\inst{4},
C. Gry\inst{6}, J. Guibert\inst{7}, O. Moreau\inst{7} and F. Tajhmady\inst{7}\\
}
\institute{
Harvard-Smithsonian Center for Astrophysics, 60 Garden St., Cambridge, MA 02138, USA.
\and
Institut d'Astrophysique de Paris CNRS, 98bis boulevard Arago 74014 Paris France. 
\and
CEA, DSM/DAPNIA, Centre d'\'etudes de Saclay, 91191 Gif-sur-Yvette, France.
\and
Laboratoire de l'Acc\'el\'erateur Lin\'eaire IN2P3, Centre d'Orsay, 91405 Orsay, France.
\and
Observatoire de Marseille, 2 place Le Verrier, 13248 Marseille 04, France.
\and
Laboratoire d'Astronomie Spatiale CNRS, Travers\'ee du siphon, les
trois lucs, 13120 Marseille, France.
\and
Centre d'Analyse des Images de l'Institut National des Sciences de l'Univers, CNRS
 Observatoire de Paris, 61 Avenue de l'Observatoire, 75014 Paris, France.
}
\offprints{D. Sasselov (USA address)}

\date{Received;accepted}
\maketitle 
\markboth{D.D.Sasselov,J.P.Beaulieu,C.Renault, et al.}{D.D.Sasselov,J.P.Beaulieu,C.Renault, et al.}

\begin{abstract}
This is an investigation of the period-luminosity relation of classical
Cepheids in samples of different metallicity. It is based on 481 Cepheids
in the Large and Small Magellanic Clouds from the blue and red filter 
CCD observations (most similar to $V_J$ \& $R_J$) of the French EROS
microlensing project. The data-set is complete and
provides an excellent basis for a differential analysis between LMC and SMC.
In comparison to previous studies of effects on the PL-relation, the EROS
data-set offers extremely well-sampled light curves and well-filled 
instability strips. This allows reliable separation of Cepheids pulsating
in the fundamental and the first overtone mode and derivation of differential
reddening.

Our main result concerns the determination of distances to galaxies which
are inferred by using
the LMC as a base and using two color photometry to establish the amount of
reddening. We find a zero-point offset between SMC and LMC which amounts to
a difference between inferred and true distance modulus of 0.14$\pm 0.06$ 
mag in the $VI_c$ system. The
offset is exactly the same in both sets of PL-relations $-$ of the
fundamental and of the first overtone mode Cepheids. No effect is seen
on the slopes of the PL-relations, although the fundamental and the
first overtone mode Cepheids have different PL slopes. We attribute the
color and the zero-point 
offset to the difference in metallicity between the SMC and LMC Cepheids.
A metallicity effect of that small magnitude still has important
consequencies for the inferred Cepheid distances and the determination of
$H_0$. When applied to recent estimates based on $HST$ Cepheid observations,
our metallicity dependence makes the low-$H_0$ values
(Sandage \etal 1994) $higher$
and the high-$H_0$ values (Freedman \etal 1994b) $lower$,
thus bringing those discrepant estimates into agreement near
$H_0 \sim 70$ \kms $Mpc^{-1}$.
\\
\thanks{*~This work is based on observations at the European Southern Observatory,
La Silla, Chile.}
\end{abstract}
\section{Introduction}
\par
The \ce period-luminosity (PL) relation is widely accepted as being one of the
most accurate primary distance indicators to nearby galaxies. It has now been
applied as far as the Virgo cluster of galaxies (Freedman \etal 1994b,
Pierce \etal 1994, Sandage \etal 1994, Tanvir \etal 1995) and holds the
greatest promise to settle the debate over the value of the Hubble constant
by providing a $\pm10$\% accurate determination (the goal of the $HST$ Key
Project).

The Cepheid variable stars offer a simple way of measuring distances: their
pulsation periods (easy to obtain) are strongly correlated with their
luminosities; then distances are determined from their apparent brightness
via the PL relation. The reliability of these distances stems from the good 
theoretical understanding of the Cepheids and their PL relation (Iben \& Tuggle
1975). Theory predicts a small, but not negligible, abundance effect on the
PL zero point (see Stothers 1988 for review). However the ambiguous results of 
previous empirical tests for this effect have led all recent $HST$ studies
quoted above to assume that the \ce PL relation is insensitive to metallicity.

Three sources contribute to an abundance dependence of the PL relation:
(1) theory of stellar pulsation, through the dependence of period on mass and
radius; (2) theory of stellar evolution, through the mass-luminosity
relation; and (3) theory of stellar atmospheres, through line blanketing and
backwarming, $i.e.$ the relations between effective temperature, absolute
magnitudes in bandpasses, and bolometric correction. 
As a result, a metal-poor Cepheid is always fainter than a
metal-rich Cepheid (at a fixed period and temperature); metal-poor Cepheids
are also hotter (bluer) on the average. Given the sensitivity of each of
the above three sources to metallicity, the overall weakness of the effect
on the bolometric PL relation is remarkable (Stothers 1988, Chiosi \etal 1993).
Theory predicts that the slope of the PL relation is nearly independent of
metallicity, only the zero point is affected. The observed PL relations are
not bolometric, hence the effect would depend on the bandpass. Most of the
above predictions are valid for a limited range of Cepheid pulsation periods 
$-$ Cepheids with P$\geq$50 days are expected (and observed) to deviate from
a linear PL relation.

Theory in itself predicts a period-luminosity-color relation (Sandage 1958).
However we study the PL relation, because this is the current one of choice
in the determination of $H_0$. Theory predicts abundance effects on the
PL relation due to helium (Y) and heavy elements (Z). However we only know
the heavy elements abundances (Z) in Magellanic Clouds \cs, hence we speak
of metallicity effects in this paper.

Previous studies have looked for metallicity effects on the Cepheid PL
relation since the beginning of the 70s, but the issue remains unsettled
observationally. Partly to blame is the near degeneracy between
three important observed properties of Cepheids: the lines of constant
period, reddening, and metallicity, in optical and near-infrared bandpasses.
The lines of constant 
period represent the range of temperatures over which a star can sustain
a stable Cepheid pulsation in a given mode and period; this introduces
a natural width (in luminosity) to the PL relation. The amount of obscuration
(reddening) is a more serious problem, as it is common practice to derive
it from the photometry of the Cepheids themselves $-$ thus the color difference
due to metallicity will affect the reddening estimate and the inferred
distance (Stothers 1988, Freedman \& Madore 1990).  

The first large scale comparison between SMC, LMC, and Galaxy \cs 
is due to Payne-Gaposchkin \& Gaposchkin (1973). Payne-Gaposchkin (1974)
concluded that there was no evidence for composition differences.
However, Gascoigne (1974) reinvestigated the apparent color differences
between LMC and SMC \cs as a metallicity effect and found that
the SMC Cepheids would be fainter than LMC \cs by 0.1 mag (in $V$). 
The color shift between LMC and SMC \cs was confirmed by Martin, Warren, \&
Feast (1979) and clearly distinguished from differential extinction.
Subsequently,
Iben \& Tuggle (1975) and Iben \& Renzini (1984) argued for a much smaller
effect, mostly from theoretical considerations. Stothers (1988)
offered a critical review of all these attempts and pointed out the
effect of the reddening correction. In a different approach to the problem,
Caldwell \& Coulson (1985, 1986) merged theoretical and empirical relations
and individual reddenings from color-color diagrams to derive PL relations
$adjusted$ for abundance differences. These fit well their data-set of about
130 Cepheids in LMC and SMC. Caldwell \& Coulson's results confirm the
existence of a metallicity effect. Their very different approach provides
no base for a detailed comparison with the current extragalactic use (Stothers
1988, Madore \& Freedman 1991). In addition, Caldwell \& Coulson's sample 
of about 130 \cs appears to be too small for quantifying the tiny effect.
This can be seen in a recent comparison of primary distance
indicators to 15 galactic and extragalactic objects in which
Caldwell \& Coulson's PL relations ($adjusted$ for metallicity) were used
(Huterer, Sasselov, \& Schechter 1995) $-$ the uncertainties within and
between the four primary indicators are larger than the weak metallicity
dependence. 

The main reason extragalactic distance measurements have assumed that
Cepheid luminosity does not depend on metallicity is the work by Freedman
\& Madore (1990). They found that any distance differences in three M~31
fields with varying metallicity are consistent with statistical noise,
and much smaller than Stothers'(1988) prediction. The conclusion is based
on a sample of 38 Cepheids and 152 $BVRI$ observations; the method used is
the same used in all recent distance determinations.
The situation was reviewed by Feast (1991), who urged for the
need to test the metallicity corrections empirically. This was partially
accomplished by Laney \& Stobie (1994), who concluded that their $VJHK$
data on 21 Galactic and 115 MC \cs do not support the implications of
Freedman \& Madore's result that intrinsic \ce color is independent of
metallicity. At the same time Gould (1994) challenged
Freedman \& Madore's conclusion by pointing out the high degree of correlation
among the $BVRI$ measurements (treated by Freedman \& Madore as independent).
Gould obtained a PL zero-point shift similar to that of Stothers (1988) by
reanalyzing the same $BVRI$ observations of 36 \cs in M~31. However, he
showed that the data-set suffers from some systematic uncertainty, which
affects the derived size of the effect. Stift (1995) confirmed independently
Gould's conclusions using current theoretical evolutionary and atmosphere
models. Thus the issue remained unsettled. As a result all recent HST
measurements of Cepheid distances assume that no metallicity effects are
present at V and I (Freedman \etal 1994b,Sandage \etal 1994,Tanvir \etal
1995).

In this paper we use a new data-set of two-color photometry ($\sim$
3 million observations) of about 500 \cs in the LMC and SMC to derive
the dependence of the optical PL relations on metallicity. We find that
the apparent distance modulus depends on metallicity roughly as:
$0.4([Fe/H]+0.3)$. 
The dependence is weaker than some previous claims, though by no means
negligible. It is derived from a differential LMC-SMC analysis which
incorporates all correlations between \ce measurements, independently
derived metal abundances for \cs and supergiants, and known limits to
the extinction laws. The photometric database is a byproduct of the
EROS microlensing survey (Aubourg \etal 1993a, Beaulieu \etal 1995,
Beaulieu \etal 1996).

\section{The observations}
\par
The EROS (Exp\'erience de Recherche d'Objets Sombres) French collaboration 
(Aubourg \etal 1993a, 1993b, 1995), and the MACHO project
(Alcock \etal 1993) are both searching for baryonic dark matter in the 
galactic halo through microlensing effects on stars of the Magellanic Clouds.
The EROS CCD equipment has been described by Arnaud \etal (1994). The 
observation, reduction, and calibration procedures have been described by
Grison \etal (1995). All details on the \cs discovered and analyzed in
the LMC and SMC are given in Beaulieu \etal (1995) and Beaulieu \etal (1996),
respectively.

In brief, all observations were obtained at ESO, La Silla using a 0.4-m,
f/10 reflector and a 2$\times$8 mosaic of 16 CCDs. Two broad-band filters
were used $-$ $B_E$ and $R_E$, with central wavelengths which fall roughly
between Johnson $B$ and $V$ and $R$ and $I$, respectively. The observations
were obtained during the 1991-92 and 1993-95 seasons. 

Our
differential analysis relies on no external zero-point, and is thus completed
within the $B_E$,$R_E$ system. However, in applying our metallicity effect
to the $VI$ work with the HST, we have investigated our transformation
between $B_E$,$R_E$ and $V,I$. The net transmission of the $B_E$ band is
affected by the blue cutoff of the CCDs sensitivity, and is thus much
closer to Johnson $V$ than to $B$. The $R_E$ bandpass remains intermediate
between Cousins $R$ and $I$. The $B_E$ filter is broader than Johnson $V$ in
a similar way as the $HST$ F555W filter is broader than $V$.
The $V-I$ and $B_E-R_E$ colors transform well
between each other as a result: $V-I=1.02(B_{E2}-R_{E2}), \sigma=0.02 mags$.
The LMC data from the 1991-92 season was obtain with somewhat different set
of blue-red filters. 
Therefore we treated the LMC Cepheid data as two different sets
and compared each of them separately to the SMC data. The 1991-92 season LMC
data is useful because of the large number of observations. With thousands of
stars available in the Cepheids range of magnitudes and colors on
the EROS CCD frames, constructing an accurate and reliable photometric
transformation is easy. We transformed all old $B_{E1},R_{E1}$
photometry to the new $B_{E2},R_{E2}$ system by:
$$
B_{E2}-B_{E1} = -0.680\pm0.060 - 0.175\pm0.005(B_{E1}-R_{E1}),
$$
$$
R_{E2}-R_{E1} = -0.116\pm0.060 - 0.100\pm0.005(B_{E1}-R_{E1})
$$

The two sets of LMC data are very similar and produced the same result in
the final differential analysis.

The identification of Cepheids in both Clouds is based on the Fourier
components of their light curves and mean magnitudes in the expected luminosity
range. Several Cepheids were excluded from our analysis on the basis of
strong blending. These could be either actual physical binaries, or unresolved
stars along the line of sight. They are easy to identify by the form of the
pulsation loop on the color-magnitude diagram (Figure 1) and by the accompanying
deviation of their mean magnitude and color from the rest of the \ce sample.
Examining the pulsation loops shows that background blending for the LMC and
SMC Cepheids is not a problem (affects less than 10\% of them), as should be
expected for luminous stars like them.
\begin{figure}
\includegraphics{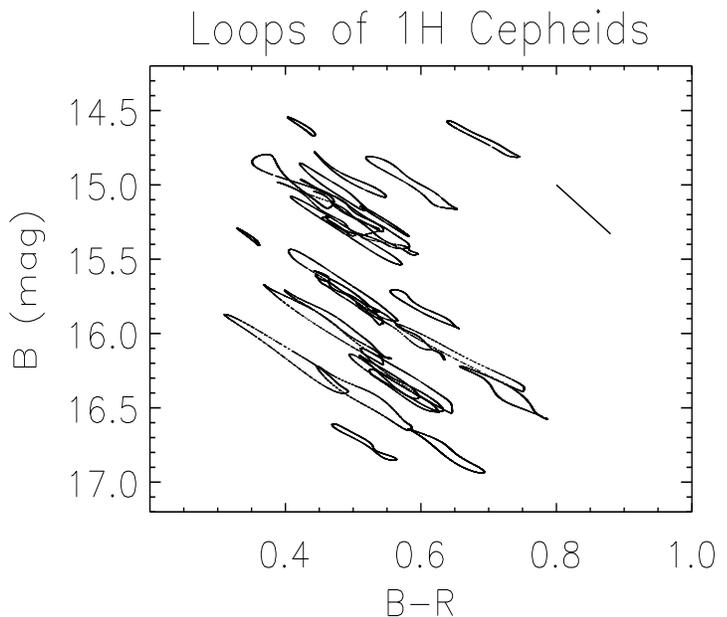}
\vspace*{3.2in}
\caption[]{The loops which Cepheids complete during a pulsation cycle in the 
temperature-luminosity plane. The form of a loop reflects the thermodynamic
relation between stellar radius and temperature change, plus a small
nonlinear effect on the emergent radiation due to the Cepheid atmosphere.
Whenever a constant flux is added to that of the Cepheid, the loop looks like
an ``eight" (a bluer blend) or has a ``histeresis"-like form (a redder blend).
One of each
examples are shown on this figure of first overtone LMC \cs $-$ in the upper
left and lower right sides of the instability strip, respectively.}
\end{figure}
We investigated several possible sources of difference between the
LMC and SMC photometry, like airmass and the Pinatubo eruption, but found
their effect to be insignificant.

\section{The concept}
\par
The concept is simple $-$ we have two complete \ce samples with known difference
in metallicity $\Delta [Fe/H]_{LMC-SMC}=0.35$ (Spite \& Spite 1991;
Luck \& Lambert 1992). This is an average difference between spectroscopic
abundances of longer-period Cepheids.
We compare the two samples in the period-magnitude-color
(PLC) manifold to derive 2 independent sources of difference $-$ distance and
extinction, and to search for a $third$ source $-$ metallicity.

For each \ce we have 3 observed quantities:
two coordinates and a pulsation period; and 2 observed data-sets: complete
light curves in $B_E$ and $R_E$. The 2 data-sets (light curves) give
two observed quantities: the time-averaged, intensity-weighted mean
magnitudes; thus each \ce has 5 observed quantities in total.
The light curves also provide independently the means to separate \cs of
different pulsation modes and to treat background blending.

We construct the PL relations in each band, $B_E$,$R_E$, of LMC and
SMC \cs, $i.e.$ the observed mean magnitudes, $Q_{i,k}$, for each \ce of period
$P_k$ and each band,$i$, are fit to a linear function:
\begin{equation}
Q_{i,k} \sim {\alpha}_i + {\beta}_i{\rm log}P_k,
\end{equation}
where ${\alpha}_i$ are the zero points and ${\beta}_i$ are the slopes;
within our completeness $-$ $P\leq 30$days, the PL relations are found to be 
linear. We compare the two \ce sets in the PL plane. They are offset in
luminosity ($Q_i$) due to two sources: difference in distance between LMC
and SMC, and difference in extinction towards LMC and SMC. We search for a
$third$ source $-$ a term due to the difference in metallicity.

Two PL relations (PL-$B_E$ and PL-$R_E$) in each of the
Clouds can give us the distance difference and the reddening difference, but
the metallicity effect remains degenerate. We search
to lift this degeneracy by going to the
period-magnitude-color (PLC) space, and imposing constraints on the
interstellar extinction. We fix the mean reddening in LMC as E(B-V)=0.10,
the value adopted by all $HST$ teams, and we adopt independent estimates for
the foreground reddenings to LMC and SMC. Now, given a large observed sample
(in a statistical sense), we can compare differentially the PLC distributions
of LMC and SMC Cepheids. Such a comparison will constitute a fit to a 
multi-parameter model. Two of the model parameters are metallicity terms.

We chose as our goal to evaluate the metallicity effects on the LMC-based
technique used by all $HST$ teams ($e.g.$ Madore \& Freedman 1991; hereafter
$-$ the modern technique) for two reasons. First, the current Cepheid-based
$H_0$ is exclusively derived by it. Second, the EROS Cepheid sample provides
an excellent opportunity to accomplish that goal, being obtained with two
filters ($B_E$,$R_E$) similar to the $HST$ F555W, F814W filters, and
being LMC-based too.

We emphasize the need to account for the correlations between magnitude 
measurements, which has not been done in its applications to-date.
This rigorous approach was pioneered by Gould (1994); our analysis draws
heavily on it and expands it. Due to the strong correlations and
near-degeneracies present in the problem, we see no alternative to the rigorous
approach.

Hereafter the notation $\delta x$ will refer to the metallicity dependence
of the quantity $x$ in the sense: 
$\Delta x_{\rm inferred} + \delta x = \Delta x_{\rm true}$, where
$\Delta x = x_{SMC}-x_{LMC}$.

\section{The method}
\par
Our method applies the same basic techniques for modeling of data (Press \etal
1994, \S 15) introduced earlier by Gould (1994), but deviates substantially
from Gould's analysis. The difference is that we solve for a metallicity
effect which is a function of bandpass/wavelength. Therefore we model 
our data in the three-dimensional PLC
manifold, as opposed to just the PL plane. 
Thus our analysis follows the current
state of theoretical understanding (Stothers 1988; Chiosi \etal 1993; Stift
1995), and takes advantage of
the fact that our SMC data-set fills densely the PLC manifold.

Here we present a simultaneous fit of the LMC and SMC data. The two ($B_E$,
$R_E$) magnitude measurements of each \ce are treated as correlated with each
other, but not with the measurements of any other \ce. Then we can write the
${\chi}^2$ merit function as
\begin{equation}
{\chi}^2=\sum_{n=1}^4\sum_{k=1}^{N(n)}\sum_{i,j=1}^2 b_{ij}^n X_{i,k}^n X_{j,k}^n
\end{equation}
where $n=1,2$ correspond to the LMC data (with $N(n)$ Cepheids) on the
period-magnitude and color-magnitude planes, respectively; 
$n=3,4$ correspond similarly to the SMC data (with $N(n)$ Cepheids);
$i=1,2$ are the two bandpasses ($B_E$,$R_E$), and where the residuals
are defined as follows:
\begin{equation}
X_{i,k}^1=Q_{i,k}-(\alpha_i+\beta_ilogP_k),
\end{equation}
\begin{equation}
X_{i,k}^2=Q_{i,k}-[a_i+b_i(Q_{1,k}-Q_{2,k})],
\end{equation}
\begin{equation}
X_{i,p}^3=Q_{i,p}-(\alpha_i+\beta_ilogP_p+\gamma_1+\gamma_2R_i+\gamma_3^i),
\end{equation}
$$
X_{i,p}^4=Q_{i,p}-[a_i+b_i(Q_{1,p}-Q_{2,p}+\gamma_2(R_2-R_1)~~~~~~~~~~~~~~~~
$$
\begin{equation}
~~~~~~~~~~~~~~~~~~~+\gamma_3^1-\gamma_3^2)+\gamma_1+\gamma_2R_i+\gamma_3^i],
\end{equation}
with $k,p=1,...,N(n)$ being the number of Cepheids in LMC and SMC, respectively.
The covariance matrices of the data are then
\begin{equation}
c_{ij}^n=\frac{1}{N(n)-1}\sum_{k=1}^{N(n)}X_{i,k}^n X_{j,k}^n
\end{equation}
for $n=1,2$, and
\begin{equation}
c_{ij}^n=\frac{1}{N(n)-1}\sum_{p=1}^{N(n)}(X_{i,p}^n -\overline{X_{i}^n})
(X_{j,p}^n -\overline{X_{j}^n})
\end{equation}
for $n=3,4$, where
\begin{equation}
\overline{X_{i}^n}=\frac{1}{N(n)}\sum_{p=1}^{N(n)}X_{i,p}^n.
\end{equation}
In equation (2), $b_{ij}$ is the inverse covariance matrix of the data,
$b=c^{-1}$.

The above set of equations illustrates our differential analysis of
LMC and SMC \cs as {\em ensembles} of stars. Each \ce with
a period $P_k$, and mean magnitudes $Q_{i,k}$ is fitted
(for each band) to a linear function of the period
$-$ the PL relation ($\alpha_i,\beta_i$), and to
the instability strip ($a_i,b_i$).
Thus, for the LMC \cs we have the residuals
in equations (3) and (4). The residuals of the SMC \cs will contain four
additional linear terms: (1) one due to the distance difference 
($\gamma_1=\mu_{SMC}-\mu_{LMC}$);
(2) one due to extinction $-$ the relative reddening $\gamma_2={\Delta}E(B-V)$, 
and $R_i$ being the
adopted reddening vector (see \S 6 below); and (3) two (one for each band) due to 
metallicity difference ($\gamma_3^i$). All of these terms simply add on to
the zero point of the PL relation (for each band), and that is what
equation (5) is all about.

The SMC residuals from equation (6) have a slight complication, which will
become clear here. 
Let $M_i$ ($i=1,2$) be the absolute magnitudes in $B_E$ and $R_E$, 
respectively. They define the PL relations for each band: 
$M_i(PL)={\alpha}_i + {\beta}_i{\rm log}P$. Then following the metallicity
term, $\gamma_3$, introduced above, the resulting changes in $M_i$ are simply:
$\delta M_1=\gamma_3^1$ and $\delta M_2=\gamma_3^2$.
Color is defined as: $(Q_1-Q_2)_0=M_1-M_2$, where the subscript ``0" means
corrected for extinction; hence the color change at a fixed period due to
the metallicity difference between LMC and SMC will be: 
\begin{equation}
\delta(Q_1-Q_2)_0=(\gamma_3^1-\gamma_3^2).
\end{equation}
This term, $\gamma_3^1-\gamma_3^2$, will appear in the period-color plane
together with the reddening $(\gamma_2)$-term alone. The equation of the
linear period-color relation is not added to equations (3)-(6) to
minimize redundancy; its slope will be simply $(\beta_1-\beta_2)$.

Now we can describe the meaning of equation (6) and the residuals in the
color-magnitude plane. 
In the color-magnitude plane we have the more complex case, where both the
color and the magnitude are affected by both the reddening $and$ the
metallicity. While the terms for the magnitude remain the same as in
equation (5), the terms for the color will come from equation (10) for the
metallicity, and from the definition of extinction color correction. The
latter definition is: $(Q_1-Q_2)_0=(Q_1-A_1)-(Q_2-A_2)$, where the $A_i$ are
the total extinctions in each band, $A_i=E_{B-V}R_i$; hence
$(Q_1-Q_2)_0=(Q_1-Q_2)-E_{B-V}(R_1-R_2)$. Thus equation (6) describes the
following aspect of our model: more extinction makes \cs redder and fainter.
As far as the metallicity terms are concerned, they are not constrained
(although theory predicts that more metals make \cs redder and brighter).
We constrain the reddening term by not allowing unphysical negative values,
and further by considering the foreground extinction towards the SMC
as its lower limit (see \S 5 and 6).

To obtain the true distance modulus $(m-M)_0\equiv{\mu}$, the observed mean
magnitude must be first corrected for extinction: 
$\mu=(Q_1-A_1)-M_1=(Q_2-A_2)-M_2$. If $A_i$ are derived from the observations
of the \cs themselves, the metallicity dependence of the true distance 
modulus will be a combination of the dependencies of the absolute magnitudes
and the color (Stothers 1988). Using the standard definitions for the
ratio of total to selective extinction, we have:
$R_i=A_i/E_{B-V}=(R_1-R_2)A_i/E_{Q_1-Q_2},$
where $E_{Q_1-Q_2}=(Q_1-Q_2)-(Q_1-Q_2)_0$. Keeping $R_i$ fixed, we obtain for
the change of distance modulus with metallicity:
$$
\delta\mu=-\delta{M_i}+\frac{R_i}{R_1-R_2} \delta(Q_1-Q_2)_0 =~~~~~~~~~~~~~~
$$
\begin{equation}
~~~~~~~~~~~~~=-\gamma_3^i+(\gamma_3^1-\gamma_3^2)\frac{R_i}{R_1-R_2},
\end{equation}
which applies to the use of a PL relation and is in units of stellar magnitude
per $\Delta$Z of metals by mass. Stothers (1988) used this case (when
$A_i$ are derived from the \cs themselves) to derive the prototype of equation
(11) from theoretical considerations. We follow the same case (referred to as
the modern technique), but derive the correction, $\delta\mu$, empirically
from the differential analysis of the magnitudes, colors, and periods of LMC
and SMC Cepheids. Therefore the application of our method (see next section) is
restricted to this specific case. We derive the metallicity correction as
$\delta\mu= \Delta \mu_{\rm true} - \Delta \mu_{\rm inferred}$.

We want to emphasize again that our fit is in the
PLC space for the sole purpose of finding the residual color difference
between the LMC and SMC \cs; we are not interested in deriving or using
a PLC relation. It is our way of lifting the degeneracy between the effects
of reddening and metallicity in the PL planes in our differential analysis
of LMC and SMC.
Freedman \& Madore (1990) used a different approach to break this degeneracy 
$-$ by setting $\gamma_1=0$, $i.e.$ by observing three fields at the same
distance but different metallicity. They cannot use our method,
because their data-set (36 \cs) is too small to fill densely the PLC
space. In other words, in the language of $\chi^2$ statistics, 
their number of data points (36) will be
comparable to the number of model parameters.

\section{The application}
\par
As described by equations (3)-(6) of the previous section, our model of
the LMC-SMC comparison has 12 model parameters. 
To solve for all 12 parameters we adopt the following external constraints:
(1) mean LMC extinction E(B-V)=0.10 as used by the HST teams; (2) foreground 
extinctions of 0.06 for LMC and 0.05 for SMC (Bessell 1991); (3) no depth
dispersion in the EROS LMC sample; and (4) the line of nodes for SMC from 
Caldwell \& Coulson (1986) to derive the EROS SMC sample depth dispersion.
These constraints are discussed in the next two sections.
Obviously, there is one more projection $-$ on the period-color plane,
which is not included in equations (3)-(6), but is trivially
derived from them. One estimates the model
parameters by minimizing in a statistical sense the residuals, given
independent constraints on some of the parameters. To
find the best fit one differentiates $\partial{\chi^2}/\partial{A_l}=0$,
where $A_l$ is the 12 element vector of model parameters:
\begin{equation}
A=(\alpha_i,\beta_i,a_i,b_i,\gamma_1,\gamma_2,\gamma_3^i) ~~~~~~~~~~(i=1,2).
\end{equation}
The $\chi^2$ merit function of equation (2) can be rewritten in a more
explicit form for our multidimensional fit:
$$
\chi^2=\sum_{n=1}^4\sum_{k=1}^{N(n)}\sum_{i,j=1}^2 b_{ij}^n
[Q_{i,k}^n-\sum_{l=1}^{12}A_lf_{i,l}^n(P_k,(Q_1-Q_2)_k)]
$$
\begin{equation}
~~~~~~~~~~~~~~~~~~~{\times}[Q_{j,k}^n-\sum_{l=1}^{12}A_lf_{j,l}^n(P_k,(Q_1-Q_2)_k)]
\end{equation}
The differentiation yields a matrix equation which can be solved
for $A_l$:
\begin{equation}
A_l=\sum_{q=1}^{12}C_{lq}D_q,
\end{equation}
where $C_{lq}$ is the covariance matrix of the model parameters vector,
\begin{equation}
C^{-1}\equiv{B}=\sum_{n=1}^4\sum_{k=1}^{N(n)}\sum_{i,j=1}^2 b_{ij}^n
f_{i,k,l}^nf_{j,k,q}^n,
\end{equation}
with $f_{i,k,l}^n$ being the basis functions from equation (13), 
given explicitly in
equations (3)-(6); in a similar fashion, the vector $D_q$ relates the basis
functions to the mean magnitudes $Q_{i,k}^n$.

The procedure requires an iteration (see Gould 1994). First, we
construct period-magnitude and color-magnitude
relations for the \cs in the LMC and derive the eight parameters
$\alpha_i,\beta_i,a_i,b_i$ from linear fits to the data. These are then used
to find the covariance matrices of the LMC data (equation 7). Next, we
compute the covariance matrices of the SMC data and
find the best-fit values. We found that the slopes of the PL relations
do not differ (within the uncertainties) between LMC and SMC; we adopt
two unique slopes (one for fundamental, one for first overtone \cs) for both
LMC and SMC. At this point we derive $\Delta \mu_{\rm inferred}$, by
setting $\gamma_3^1=\gamma_3^2=0$. Then we minimize the residuals along
a given reddening vector R$_V$ and additionally constraining $\gamma_2$
so that SMC \cs have no unphysical extinctions (negative).
This constraint is derived independently of the fit (see next section).
We use the same form of the Galactic extinction law (Cardelli, Clayton, \&
Mathis 1989) for all samples.
We iterate this procedure until the best-fit parameters agree
with the trial parameters used to estimate the covariances.
Note that the entire procedure is separately done for the
fundamental mode and first overtone mode \cs, and thus we have a
strong constraint on the final derived parameters, which
are invariant to the type of Cepheids used to derive them.

For illustration, the covariance matrix of the LMC fundamental \cs (equation
7) is:
\begin{equation}
c_{ij}^1=\left( \begin{array}{c}
0.048~~~0.041 \\ 0.041~~~0.037
\end{array}  \right)
\end{equation}
and for the SMC fundamental \cs (equation 8):
\begin{equation}
c_{ij}^3=\left( \begin{array}{c}
0.103~~~0.073 \\ 0.073~~~0.084
\end{array}  \right).
\end{equation}
The correlation coefficients are obviously very high, as also seen in
Figs.5 \& 6.

For Cepheids all at the same distance and selected only by magnitude
the PL relation is subjected to bias, which can be avoided by using
the inverse regression (Schechter 1980, Feast 1995). We do not expect strong
magnitude bias in our samples, but used both direct and inverse
regressions in our fits.

The method above contains as a subset the procedures which comprise the
modern technique for determining \ce distances (Madore \& Freedman 1991).
Namely, one begins by 
constructing the  PL relations in the LMC, then fitting the mean magnitudes
of the SMC \cs to the LMC in each band, and deriving apparent distance
moduli for each band. These distance moduli are then fitted to an
extinction law with an adopted parameter ($R_V$) to derive the amount
of reddening in the SMC data. Finally the reddening corrected, true
distance modulus is determined. As brief as it is in this description,
the technique overlooks several sources of physical scatter and systematics
which could be present in the data.  We study those below, since they are
also part of our method to derive the metallicity dependence.
They provide the independent constraints, which make possible the derivation
of both $\gamma_3^1$ and $\gamma_3^2$.

\section{The Extinction Law and Reddening}
\par
In our differential analysis between LMC and SMC the effect of interstellar
reddening (and obscuration) is represented by the term $\gamma_2 R_i$,
where $\gamma_2=\Delta E(B-V)$ and the reddening vector
\begin{equation}
R_i=(3.81,2.48).
\end{equation}
We use the $R_V$-dependent Galactic extinction law of Cardelli, Clayton, \&
Mathis (1989) and the above values correspond to a convolution of the
EROS filters transmission curves with the $R_V=3.3$ extinction law (Fig.2);
here $R_V=A(V)/E(B-V)$.
\begin{figure}
\includegraphics{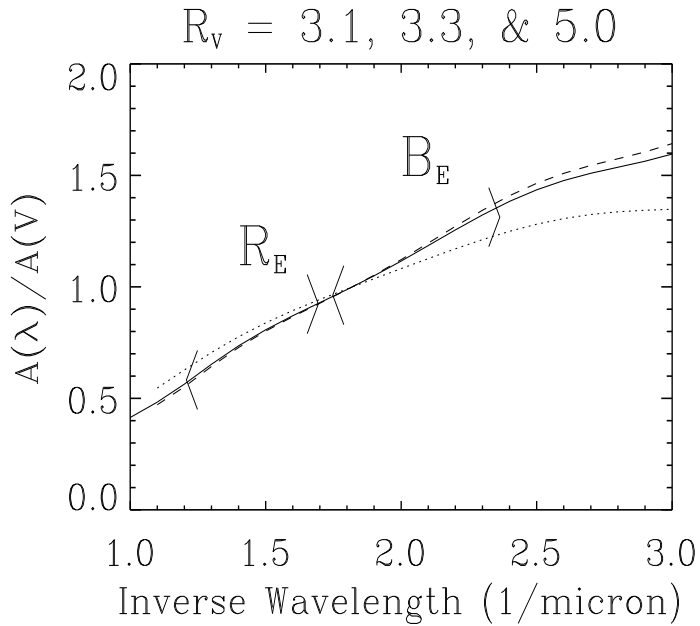}
\vspace*{3.2in}
\caption[]{The extinction diagram for three different extinction laws $-$
with $R_V$=3.1 (dashed), $R_V$=3.3 (solid), and $R_V$=5.0 (dotted). Here $A(V)$
corresponds to the amount of extinction in the Johnson $V$ band. The two
EROS filters are marked on the diagram.}
\end{figure}
An extinction law with $R_V=3.3$ is widely accepted for the extinction in
LMC ($e.g.$ Freedman \etal 1994a), while for our Galaxy $R_V=3.1$ is more 
common on the average. There are a few lines-of-sight with $R_V$ as high
as 5.3, but that's very rare. Physically, high values of $R_V$ appear to
be related to systematically larger dust particles in dense regions. However,
an empirical relation between the spectral ``bump" at 2200~\AA ~and $R_V$
seems to lack physical grounds (Cardelli, Clayton, \& Mathis 1989). When
used, this relation predicts a high $R_V$ value (as high as 5) for the
SMC, where the 2200~\AA ~feature is known to be very weak (Bessell 1991).

This is a good enough reason for us to investigate how our results are
affected by the value of the extinction law parameter. Therefore we have
convolved the transmission curves of the EROS filters with $R_V=3.1$ and
$R_V=5.0$ extinction laws as well (Fig. 2). After deriving the apparent
distance moduli between SMC and LMC for each band $and$ each mode of pulsation,
we plot them against inverse wavelength at the effective centers of the
photometric bands (Fig.3). We fit through the points Galactic extinction laws 
with different amount of extinction, as measured by E(B-V), and with
different parameters, $R_V$.
\begin{figure}
\includegraphics{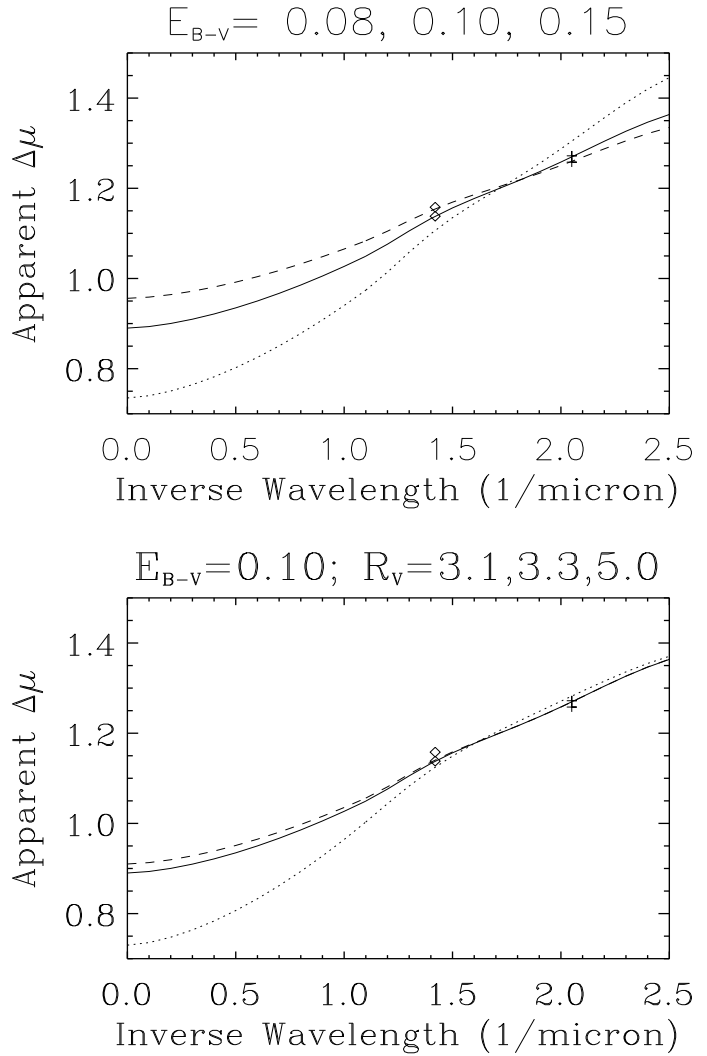}
\vspace*{5.2in}
\caption[]{The differential apparent distance moduli for SMC \cs fitted
with Galactic extinction laws: (a) for three different amounts of extinction,
and (b) for three different reddening parameters, $R_V$. At each band
we have two independent values for the distance moduli $-$ for fundamental
and firts overtone \cs, respectively. The error bars on the values are
as big as the symbols. No correction for metallicity effects has been
applied at this point. The best fit for SMC is E(B-V)=0.09 and $R_V$=3.3,
but higher values of $R_V$ are not excluded.} 
\end{figure}

Concerning the reddening of the Magellanic Clouds, it is important for our
study to know the existing observational limits for the average values.
First of all, we have the limit of foreground reddening. Towards the SMC
the foreground (due mostly to our Galaxy)
is smooth and E(B-V) lies between 0.04 and 0.06 mag (Bessell
1991). Within the SMC the reddening could vary between 0.06 and 0.3 mag.
On the average the reddening of the LMC is very similar: foreground of
0.04 to 0.09, and same within (Bessell 1991).
These values are derived independent of Cepheid observations and will prove
very useful in our derivation of a metallicity dependence.
Also, when we refer to \cs in LMC and SMC, one should bear in mind that the
\ce samples are located in the bar of LMC and near the central region of
SMC. This is particularly important in terms of their mean reddenings
(as samples); the central region of SMC is known to be dusty.

The extinction diagram in Fig.3 provides an opportunity to derive 
individual reddenings for each \ce on our list, by fitting an extinction law
with a fixed reddening parameter. The individual \ce reddenings are excellent
for the study of the structure of the PL relation (see next section).
In deriving individual reddenings we required that
no \ce be assigned a reddening smaller than the known foreground reddening
to the MC given above. As it will become clear from the next section, the
so derived individual reddenings offer no additional advantage in the 
differential analysis.

It is well known that interstellar dust absorbs or scatters light in the
optical range, hence extinction is by definition a non-negative quantity.
In our estimate of the extinction, $A_i$, we need to bear this cutoff in
mind, because $A_i$ is small towards the LMC and SMC, and for a large
subsample of our \cs we get negative values. Given the crucial use of the
reddening cutoff in our analysis and final fit to the model, it is
important to truncate the measurements of $A_i$ properly, and preserve
a useful estimate of the measurment uncertainty, $\sigma_{A_i}$ in the
process of doing so. The solution to finding the value of $A_i$ and its
error $together$ with prior knowledge that $A_i$ cannot be less than zero
is given by Bayes's theorem. For an estimate of $A_i$ with value $x$ and
normal error $\sigma_x$, and our knowledge of the distribution of the
observed $A_i$, $p(A_i)$, the probability distribution for the true $A_i$
will be given by the Bayesian filter:
$$
 p(A_i\mid x,\sigma_x)= \frac{p(x\mid A_i,\sigma_x)p(A_i)}{p(x)} =
              \frac{p(A_i) e^{-\frac{(A_i-x)^2}{2{{\sigma_x}^2}}}} 
               {\int_{0}^{\infty} p(A_i)e^{-\frac{(A_i-x)^2}{2{{\sigma_x}^2}}} dA_i}
$$
A practical, and entirely satisfactory assumption would be that $p(A_i)$ is
a one-sided gaussian which has a maximum at $A_i=0$ and declines for large $A_i$
(see Press 1996).

The conclusion of our analysis is that the reddening law in the bar of
SMC is not very different from that of the LMC with a reddening parameter
much closer to 3.3 than 5. Concerning the absolute value for the LMC
reddening (currently used: E(B-V)=0.10 and $R_V$=3.3), our 
mean value, from the 
individual reddenings, is E(B-V)=0.10$\pm$0.007.
With the derived parameters in the next section we could constrain the
mean reddening of the SMC Cepheids to E(B-V)$\geq$0.10.

\section{The PL relations - structure of the scatter}
\par
Now we are ready to apply our method. We start by constructing the PL
relations (in each band) for the LMC and SMC \cs, separately for fundamental
and first overtone pulsators (Fig.4).
\begin{figure}
\includegraphics{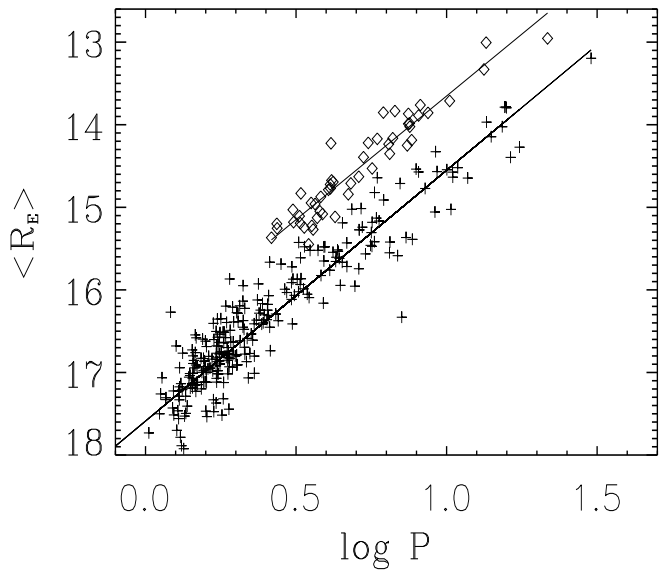}
\vspace*{3.2in}
\caption[]{The PL-$R_E$ relations for LMC and SMC \cs.}
\end{figure}
All PL relations are essentially linear in the range of periods in our data.
The slopes, $\beta_i$ (eq.1), which are insensitive to moderate amounts of 
reddening, are the same in LMC and SMC: \\
$\beta_1$=$-$2.78$\pm$0.16 and
$\beta_2$=$-$3.02$\pm$0.14 for fundamental mode \cs in LMC, and \\
$\beta_1$=$-$2.72$\pm$0.07 and $\beta_2$=$-$2.96$\pm$0.06 for SMC.
Similarly, for the first overtone \cs we have:  \\
$\beta_1$=$-$3.41$\pm$0.22 and $\beta_2$=$-$3.43$\pm$0.20 for LMC, and
$\beta_1$=$-$3.46$\pm$0.14 and $\beta_2$=$-$3.52$\pm$0.13 for SMC.
This lack of metallicity dependence of the slopes, as well as the difference
between $\beta_1$ and $\beta_2$ are in very good agreement with theory
(Stothers 1988).

The scatter about the PL relations results from physical reasons, as well as
statistical noise. On the basis of our very well sampled light curves in
two bands, we have already excluded Cepheids which show evidence of
blending, as well as stars which are not \cs, and can analyse the other
physical sources of scatter.

Individually, each \ce in the LMC can be offset from the PL relation
of the LMC \cs due to 3 $independent$ sources: (1) position in the
instability strip; (2) differential reddening within the LMC; (3) differential
distance (depth) within the LMC. These three additional independent sources 
of scatter contribute to a natural width (in L) of the PL relation.
They are manifested in the structure of the LMC covariance matrix (equation 7),
where none of the sources lies in the plane defined by the other two sources;
hence the covariance matrix would be three-dimensional. We illustrate this
by plotting against each other the magnitude residuals defined in equation (3).
In Figure 5 we have the diagram of the residuals for the LMC fundamental \cs.
The high level of correlation is evident. The diagram of the residuals shows
that reddeing (solid line) within the LMC is not the only source of scatter.
In Figure 6 we have the same diagram for the SMC \cs (equation 8); all of the 
above applies to them as well.
\begin{figure}
\includegraphics{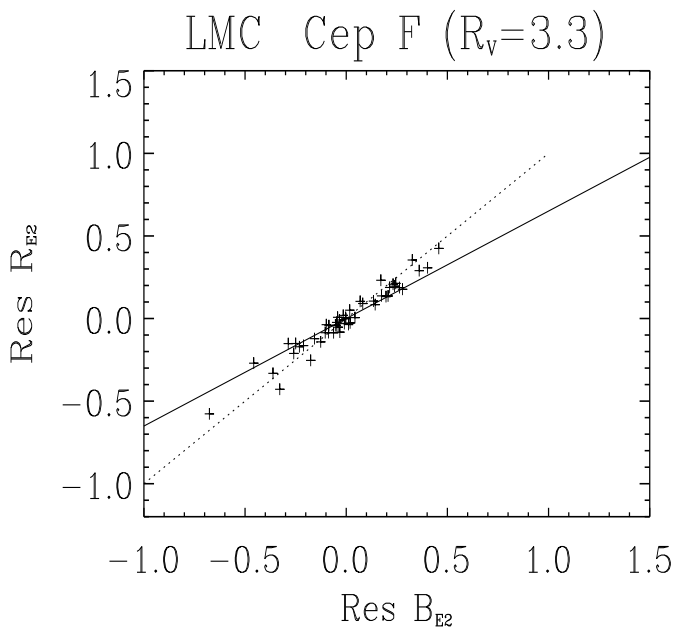}
\vspace*{3.2in}
\caption[]{The residuals in $B_E$ and $R_E$, as defined in equation (3),
plotted against each other for
the fundamental \cs in LMC and their PL relations. The solid line is the
reddening line for $R_V$=3.3. The line of constant periods is steeper.
Dispersion due to distance is even steeper $-$ along the diagonal (dotted line).}
\end{figure}
\begin{figure}
\includegraphics{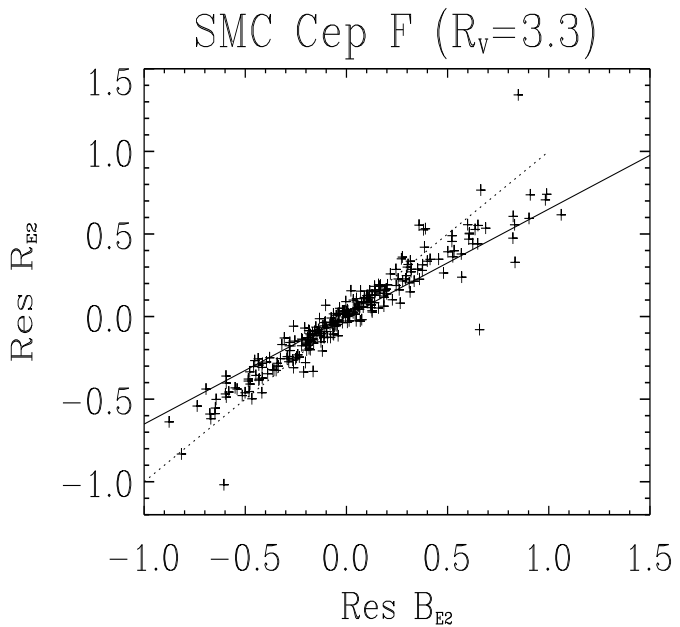}
\vspace*{3.2in}
\caption[]{The same as Fig.5, but for the fundamental \cs in SMC. The solid
line is the reddening line for $R_V$=3.3. The SMC \cs show a larger scatter,
which is also closer to the diagonal line (due to distance dispersion).}
\end{figure}

The diagrams of the residuals provide a nice illustration to the structure
of the PL relations. For example, if the PL relation had no dispersion, all
points would lie in the center (0,0) of the diagram (Fig. 5). Alternatively,
if there were groupings of \cs separated in distance along the line of sight,
they will appear clearly as clumps on the diagram.

If we deredden each \ce as described in the previous
section \S 6, we are mostly left with the dispersion due to depth (Fig. 7).
The reason is that the individual dereddening corrects also for
the width of the instability strip along the lines of constant period,
at least partially. The correction is partial for two reasons: (1) the
reddening and the $P=const.$ lines are not completely degenerate (Fig. 8);
and (2) the individual reddening is limited (from below) by the amount
of foreground reddening (independently known).
\begin{figure}
\includegraphics{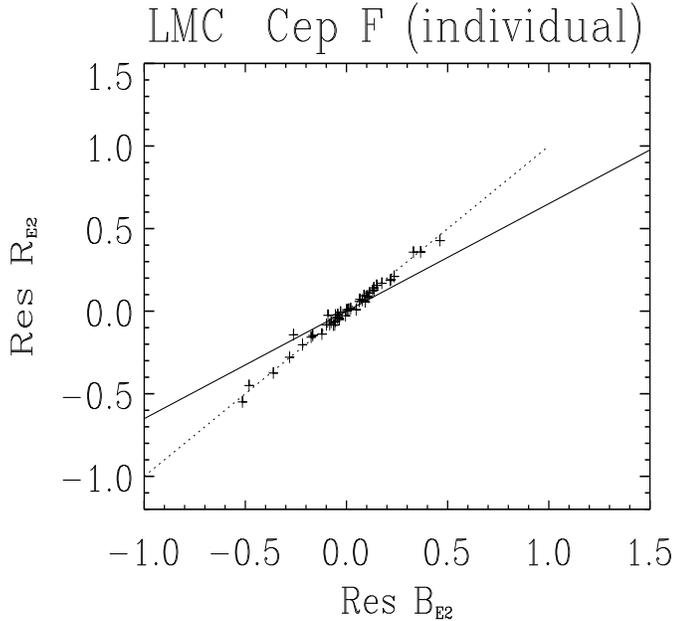}
\vspace*{3.2in}
\caption[]{The same as Fig.5, except that each \ce has been dereddened
individually using the Galactic extinction law with $R_V$=3.3. In comparison
to Fig.5, the dispersion has decreased and has shifted slope to the
diagonal. All these \cs are in the bar of the LMC, which is face on, and
that explains the very small depth dispersion we see. Conversely, in SMC
we see a substantial depth (Fig.6).}
\end{figure}
\begin{figure}
\includegraphics{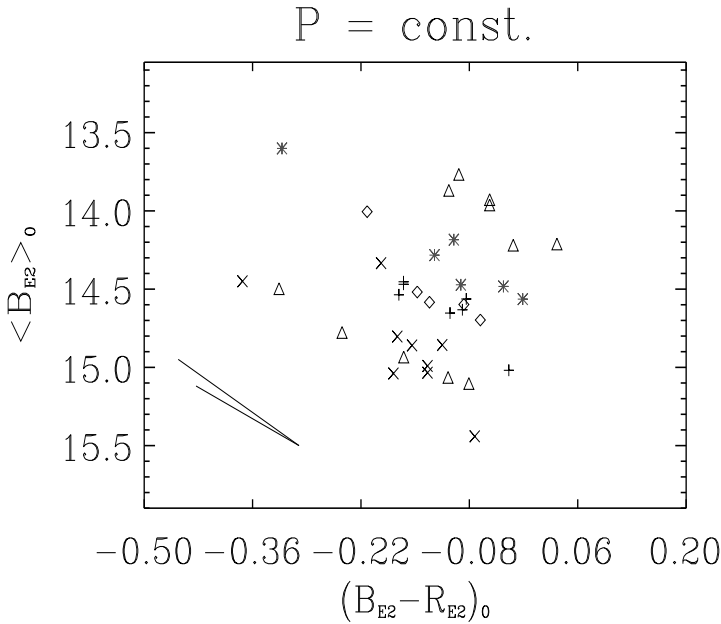}
\vspace*{3.2in}
\caption[]{An excerpt of the \ce instability strip showing groups of
\cs which happen to have the same periods $-$ from top to bottom: 7.5, 5.5,
4.2, and 3.1 days, respectively. The open symbols correspond to LMC \cs,
the rest $-$ to SMC ones. The \cs shown have been dereddened as an ensemble
in each MC, with $R_V$=3.3. The reddening lines for two values of $R_V$ are
shown in the lower left corner.}
\end{figure}

The near-degeneracy of reddening and $P=const.$ lines in the color-magnitude
plane is worth investigating. We find that the slopes of the 
$P=const.$ lines in LMC and SMC are practically the same: $3.3\pm0.4$;
while the slopes of reddening ($B_E$) are 2.86 (for $R_V$=3.3) and
3.53 (for $R_V$=5.0). A limited illustration to this is given in
Fig.8; the \cs shown have been selected to have exactly the same
periods (to the second digit) $-$ in practice we use the whole data-set
to derive the slopes. 

Now we turn to accounting for the
depth of LMC and SMC. In this we employ for the first time the last two
observed \ce quantities: the coordinates. We plot the magnitude residuals
versus angle on the sky across the SMC (1$^{\deg}$=0.0389 mag)
using the line of nodes for 
SMC from Caldwell \& Coulson (1986). They derived them over areas which
are about 25 times larger than the EROS CCD-array areas. The EROS areas cover
only the LMC bar (which is seen face on) and the SMC bar (seen edge-on) and
have dimensions of 25$^{\arcmin}\times$71$^{\arcmin}$ on the sky.
That is one reason we did not attempt our own derivation of the line of nodes of
the Clouds. In deriving the depth dispersion in the SMC we used Cepheids
along the line of nodes and orthogonal to it.
%
%

With the width of the instability strip derived from the color-magnitude
plane to be $\pm 0.22$ mags, we derive the depth dispersion in our SMC
sample to be $\pm 0.10$ mags. Our uncertainty in the value of $R_V$ limits
our ability to derive the internal amount of reddening within SMC to the
ranges already known (\S 6). However, it provides a very valuable constraint
from below on the overall reddening towards SMC as E(B-V)$\geq$0.10$\pm 0.01$;
otherwise half of the \cs will have negative extinctions, which is
unphysical. We use this value to constrain $\gamma_2$ in our multidimensional
minimization (equations 12-15).

 Finally, we would like to point out that the EROS sample of SMC Cepheids happen
to be mostly from the far arm and the main bar (see Figs.7\&9 of Caldwell
\& Coulson 1986). Hence the sample is not likely to define the centroid of
SMC well, and would result in a larger distance modulus difference with
the LMC (by about 0.15 $mags$). This is of no consequence to our
differential analysis here.

The diagrams of the residuals are independent of differential
reddening or metallicity between the MCs, however understanding the
structure of each PL relation and quantifying the sources of scatter
improve the reliability of the differential analysis.
When we $compare$ the LMC and SMC PL relations, the above three sources of
scatter are not involved (as long as we have a statistically large sample
of data points, $i.e.$ Cepheids) $-$ we use $ensemble$ quantities. Then it
is only consistent to use the ``ensemble" reddening for each MC, the
individual reddenings will introduce a bias. Other sources of scatter have
been proposed in a critique of the PLC relation $-$ mass loss and different
strip crossings (Stift 1982, 1995). We do not identify another source in
our data and in the remaining scatter; such effects must be subtle for the
PL relation alone.

\section{Results}
\subsection{The effect of metallicity}
\par
We apply the procedure in \S~5 to the data-sets of fundamental mode
and first overtone mode \cs separately and obtain very similar results. There
are enough first overtone \cs, 27 and 141 respectively, to make this
result statistically significant. Thus, the two sets of PL relations give
the same distance modulus difference (SMC-LMC) of 0.86$\pm$0.045 mag (before
corrections for reddening and metallicity). This illustrates for the
first time the usefulness of the PL relation of first overtone \cs for
determining distances. 

The model parameters derived from the fits of the different datasets
are as follows: 
$\alpha_{i}',\beta_{i}$= 17.61$\pm 0.035$,-2.72$\pm 0.07$; 17.74$\pm 0.029$,-2.95$\pm
0.06$; $a_{i}',b_{i}$= 14.49$\pm 0.13$, -11.78$\pm 0.94$; 14.36$\pm 0.12$, -12.77$\pm
0.90$; $\gamma_1$= 0.62$\pm 0.04$, $\gamma_2$= 0.010$\pm 0.01$, $\gamma_3^1$=
0.06$\pm 0.01$, and $\gamma_3^2$= -0.01$\pm 0.01$.
This is the fit with the LMC 1993-94 data; the zero points $\alpha_{i}'$,
$a_{i}'$ refer to the SMC linear fits. 
The correction due to the metallicity dependence of the inferred distance
modulus by the described technique is then $\delta \mu=\Delta \mu_{true} -
\Delta \mu_{inferred} = -0.145 \pm 0.06$mag.
If the LMC 1991-92 dataset is used, the color shift is smaller by 0.02 $mags$
to $\gamma_3^1-\gamma_3^2$= 0.05$\pm 0.01$, but a corresponding change in the
PL relations leads to $\gamma_1$= 0.63$\pm 0.04$, $\gamma_2$= 0.025$\pm 0.015$,
$\gamma_3^1$=0.05$\pm 0.01$, and $\gamma_3^2$= -0.01$\pm 0.01$.
Therefore, $\delta \mu= -0.139 \pm 0.06$mag.
In the error budget, in order of
importance, the sources are: (1) the photometric transformation between
the 1991-92 and present data (0.06 mag); (2) the scatter in the PL relations
(0.04 mag); (3) the uncertainty in the reddening (and depth) estimate in
SMC (0.03 mag). The first error source is not applicable to the LMC 1993-94
dataset.

The above result is for a fixed reddening vector, $R_i$ (equation 18).
If we change the parameter to $R_V$=5.0, $i.e.$ $R_i$=(5.51, 3.95),
the above result for the metallicity
effect is no longer unambiguous. The reason for this is that now it is possible
to satisfy the constraint from the foreground SMC reddening and correcting
for the offset of the SMC \cs in the color-magnitude plane. In other words,
if in SMC $R_V$=5.0 and E(B-V)=0.04, one need not invoke a metallicity
effect in the differential comparison to LMC. The fact that the SMC \cs
appear bluer than the LMC \cs is then explained with the small amount of
extinction and steep extinction law. 

While a value of $R_V \geq$5.0 is not excluded (but unlikely) by our data,
an extinction of E(B-V)=0.04 is very difficult to accommodate in our
analysis of the magnitude residuals (see \S 7). Our mean value is
0.125$\pm$0.009, and that is in very good agreement with a number of
independent estimates, as reviewed by Bessell (1991).

Therefore we consider this alternative explanation (very high $R_V$ and
very low reddening) as very unlikely. We propose instead that the observed
effect is due to the known difference in metallicity between the 
Cepheid samples of the LMC and SMC: 0.14 $mags$ for a factor
of 2 difference in heavy elements. The sign of the effect would lead to
the distance of a metal-poor sample to be overestimated.

In applying the above result to other galaxies, we need to address two
issues: (1) about the form of the metallicity dependence of inferred \ce 
distances, and (2) about the role of the helium abundance ($Y$).
The form of the dependence can be inferred from current theoretical
models. Photometric colors and bolometric corrections (from model
atmospheres) have linear dependencies on $log$($\frac{Z}{0.016}$), where
$Z$ is the abundance of metals (by mass) and Z=0.016 in the Galactic
Cepheids. Mass-luminosity relations, hence $-$ $M_i$, seem to depend
linearly on $Z$ (Chiosi, Wood, \& Capitanio 1993; Buchler \etal 1996). 
In addition, these models
show a significant linear dependence on the abundance of helium ($Y$).
We have currently no data on $\Delta$Y between the LMC and SMC Cepheids,
therefore in applying our metallicity dependence on distance to other
galaxies, we have to assume that 
$(\Delta{Y}/\Delta{Z})_{SMC-LMC} = (\Delta{Y}/\Delta{Z})_{galaxy-LMC}$.
That assumption is strongly supported by the current level of
understanding and measurement in stellar populations relevant to our
study (e.g. Olive, Skillman, \& Steigman 1996), and will not affect any
conclusions with regards to distances. We also want to point out that
the above needs to be born in mind in comparisons to results from
theoretical modelling, e.g. Chiosi \etal (1993) use Y=0.27 for both
LMC (Z=0.008) and SMC (Z=0.004) in Table 15, which may not be a good assumption.

In view of the above and our model,
we derive the following metallicity dependence of inferred \ce distances:
\begin{equation}
\delta {\mu}=(0.44_{-0.2}^{+0.1}) ~log\frac{Z}{Z_{LMC}},
\end{equation}
where $Z$ is the abundance of metals (by mass) in the studied \cs and 
$Z_{LMC}$=0.0085.
The metallicity dependence is valid in the spectral region covered by
the EROS filters. Due to these filters transformation properties into the
standard $BVRI$ system, and the application at hand, we have derived it
for V, and V-I use in particular, where $V-I=1.02(B_E-R_E), \sigma=0.02 mags$.
The above metallicity dependence applies to distances inferred by
differencing against the LMC \cs and Cepheid derived reddenings
(what we refer to as the {\em modern technique}),
and $R_V(LMC)$=3.3.

The metallicity dependence we find is two times less steep than that
found by Gould (1994) from Freedman \& Madore's M31 data (0.88$\pm$0.16~[Fe/H]).
It is also smaller than Stothers' prediction (28.7~$\delta{Z}$, for
$\delta{Y}/\delta{Z}=3.5$).
However our result is in perfect agreement with the metallicity effect
seen in Galactic \cs as a function of galactocentric distance. The
observed abundance gradient for Cepheids is 
$\delta{[Fe/H]}/\delta{R_{GC}}=-0.07 \pm 0.02 kpc^{-1}$ (Giridhar 1986).
Two independent \ce samples show the same effect of progressively bluer
color with increasing galactocentric distance (Caldwell \& Coulson 1987,
and Gieren \etal 1993), which converts to the following values for 
$\gamma_3^1-\gamma_3^2$ (of equation 10) in units of [Fe/H]: 
$0.29\pm 0.05$ for (B-V), and $0.20\pm 0.05$ for (V-I). For (V-I) of
our system we obtained $0.20\pm 0.02$.
Exactly the same value is obtained (and a nice illustration of the effect) from
the compilation of LMC and SMC Cepheids in the study by Di~Benedetto (1995) $-$
see his Fig.1 \& 2, and also 8.  

Our comparison to the work of Caldwell \& Coulson (1986) in $BVI$ and of
Laney \& Stobie (1994) in $VJHK$ is limited for at least two reasons. First,
we cannot reproduce their reddening determinations with only two bands.
Second, we are unable to comment on the theoretical models which they used
as constraints. Nevertheless, we find a similar intrinsic difference in
$V-I$ colors between LMC and SMC as Caldwell \& Coulson; our value is two
times larger. We also agree with Laney \& Stobie's main conclusions,
despite the uncertainty of a $J$ to $I$ transformation needed for such a
comparison.

The metallicity dependence is not strong enough to upset the overall
agreement of the four primary distance indicators in fifteen local
distances (Huterer, Sasselov, \& Schechter 1995). Similarly, the discrepant
distance moduli involving IC~1613 noted by Gould (1994) are less discrepant
with our weaker metallicity dependence; the RR Lyrae distances may be
partly responsible for the remaining difference.

\subsection{Implications for $H_0$}
Despite its relative weakness, this metallicity dependence
has a significant effect on extragalactic distance measurements.
The effect of metallicity on Cepheid luminosity, when not accounted for, 
can lead to two types of errors in distance measurements (Gould 1994):
incorrect $H_0$ value, or a dependence of $H_0$ on distance appearing as
proper motions or Malmquist bias. Here we illustrate the effect
of our metallicity dependence on the value of $H_0$, as based on \cs.

Recent efforts to determine $H_0$ from \ce distances have focused on:
(1) the Virgo and Fornax cluster galaxies (Freedman \etal 1994b, 1996),
(2) the Leo~I group, containing ellipticals (Tanvir \etal 1995); and
(3) the parent galaxies of supernovae of type Ia (Sandage \etal 1994).
All these studies use the same modern technique with the LMC as a base
and all the same initial assumptions; we share them in our analysis.
Yet they result in three different values of $H_0$, with hardly overlaping
error bars.
The abundances of metals in the young populations of the host galaxies
for these studies differ and, apparently, in a systematic way. 
Therefore we propose that the metallicity dependence may be responsible
for most of this discrepancy. To illustrate our idea, below we apply 
Equation (19); please note that the uncertainties in the metallicities
of the extragalactic Cepheids are not yet well known, hence we do not
try to provide rigorous error estimates to the corrected values.

With the first approach and HST $VI$ photometry of \cs in M100, Freedman
\etal (1994b) derive $H_0=80\pm17$ \kms $Mpc^{-1}$. For the metallicity
of the \cs we adopt [Fe/H]=+0.1 (Z=0.021), derived from the abundances of
H~II regions (Zaritsky \etal 1994), assuming a solar [O/Fe] ratio.
With the uncertainty range for the slope of our metallicity effect, this leads
to $H_0=76-70$ \kms $Mpc^{-1}$. With some of the color difference due to
metallicity effects, the estimate of interstellar reddening will also
change (the metal-rich M100 Cepheids are intrinsically ``redder" than the
LMC ones) $-$ $e.g.$ from E(V-I)=0.13$\pm0.06$ (after Ferrarese \etal 1996)
we obtain 0.05 $mags$.
Recent interim results reported on half-dozen
more galaxies (Freedman, Madore, \& Kennicutt 1996) include mostly spirals 
which are not as
metal rich as M100, {\em e.g.} M101 has LMC abundances in its outer field.
All this brings their interim result down to 73$\pm10$
from their initial M100-based value, as we would expect from our metallicity
effect. A complete analysis should be done when all data becomes available.

With the second approach and HST $VI$ photometry of \cs in M96, Tanvir \etal
(1995) derive $H_0=69\pm8$ \kms $Mpc^{-1}$. The abundance estimate for
M96 comes form Oey \& Kennicutt (1993), slightly below solar at [Fe/H]=$-$0.02.
This implies a small change in the \ce distance to M96
and a decrease in the value of $H_0$ by 3$-$5 \kms $Mpc^{-1}$.
The reddening changes from E(V-I)=0.09$\pm0.10$ to 0.03 $mags$.

With the third approach and HST $VI$ photometry of \cs in IC4182 and NGC5253,
Sandage \etal (1994) derive $H_0=55\pm8$ \kms $Mpc^{-1}$. Abundances in
NGC5253 have been measured and discussed by Pagel \etal (1992); discussion
of abundances in IC4182 is given by Saha \etal (1994) $-$ we adopt
[Fe/H]=$-$1.3 and caution on the large uncertainties. With this metallicity
the above estimate of $H_0$ should be increased by 6 to 13 \kms $Mpc^{-1}$.
The reddening estimates change from E(V-I)=$-$0.11$\pm0.07$ and 0.03$\pm0.15$,
to E(V-I)=0.09$\pm0.07$ and 0.23$\pm0.15$, respectively (these metal-poor
\cs are ``bluer" than the LMC ones).
Here we should point out that the interpretation of the light curves of SN Ia
themselves has been improved considerably by Riess, Press, \& Kirshner (1995),
leading to an additional systematic shift upwards
by 6 \kms $Mpc^{-1}$. 
This brings about the estimate to $H_0\approx 70$, which is
also in good agreement with the theoretical calibration of SN Ia
and the value $H_0=67 \pm9$ \kms $Mpc^{-1}$ by H\"oflich \& Khokhlov (1996).
The new value of $H_0=58\pm4$ \kms $Mpc^{-1}$ by Sandage et al. (1996)
includes two SNe~Ia in spiral galaxies, in which Cepheid metallicity is most
likely similar to that of the LMC. Note, however, that the two SNe $-$ 1981B
(NGC4536) and 1990N (NGC4639)
are by 0.2$-$0.4 $mags$ dimmer than the three SNs in IC4182 and
NGC5253. Therefore, the result we obtained above by accounting the effect
of metallicity on the Cepheid distances for each galaxy, remains virtually
unchanged, and so does the consistency with the results of Riess,
Press, \& Kirshner (1995).

The metallicity dependence we found from the LMC/SMC analysis brings all
the derivations of $H_0$ to good agreement. It also leads to reasonable
amounts of reddening, especially regarding the negative values obtained
otherwise in metal-poor samples.

Secondary distance indicators, like the Tully-Fisher (TF) relation, the
Surface Brightness Fluctuations (SBF) method, and the Planetary Nebulae
Luminosity Function (PNLF) method, have their zero points calibrated with
Cepheids. The TF relation is calibrated primarily in M81 (Freedman \etal 1994a) with
25 Cepheids. The metallicity of these \cs is high, [Fe/H]$\approx$0.05
(Garnett \& Shields 1987; Zaritsky \etal 1994). The SBF and PNLF zero points
come from M31, and in a similar fashion $-$ from metal-rich Cepheids
(Freedman \& Madore 1990). Given our metallicity effect, we are not surprised
to note that all these methods give a high value for $H_0$.

\section{Conclusions}
\par
We use the new EROS microlensing survey data-set of 3 million two-color
observations of about 500 \cs in the LMC and SMC to search for and derive
the dependence of the optical PL relations on metallicity.
We find that:\\
(1) The PL relations for both types of Cepheids have the same zero-point
offset (but no slope difference), which is attributed to the effect of
metallicity under a reasonable assumption about the extinction law.
As expected from theory, this effect of metal content is manifested in
a color shift of the instability strip. It amounts to about $\frac{1}{5}th$
of the strip width, and we could detect it unambiguously thanks to the
large number of \cs and extremely well sampled light curves.
\\
(2) With the known ensemble difference in metal content between LMC and SMC
\cs, we derive a linear relation between the distance modulus correction
and metallicity:
$$
\delta {\mu}=(0.44_{-0.2}^{+0.1}) ~log\frac{Z}{Z_{LMC}},
$$
It applies to distances which are inferred by using LMC as a base {\em and}
using two color $VI$ photometry of the Cepheids to establish the reddening.
The linearity of 
metallicity dependence is a good assumption, but needs
to be confirmed empirically outside the range of application (a factor of
few lower than SMC and higher than the Galaxy).
\\
(3) The first overtone \cs have PL relations which provide distances fully
consistent with the PL relations of fundamental mode \cs.

We use two color bands closely spaced in wavelength because of availability,
not their desirability; ideally one would like to use at least three bands,
one of them in the near-infrared.

Our result can be applied to the long-standing discrepancy between the
low-$H_0$ scale and the high-$H_0$ scale. The host galaxies 
on which each of these scales relies appear to have systematically different
metallicities. A simple application of our correction to several recent
derivations makes the low-$H_0$ values
(Sandage \etal 1994) $higher$
and the high-$H_0$ values (Freedman \etal 1994b) $lower$,
thus bringing those discrepant estimates into agreement near
$H_0 \sim 70$ \kms $Mpc^{-1}$.

\begin{acknowledgements}
We are grateful for the support 
given to our project by the technical staff at ESO La Silla and thank
D. Welch for his helpful suggestions.
\end{acknowledgements}


\begin{thebibliography}{}
\bibitem[1993]{Alc}  Alcock C. , Akerlof C.W., Allsman R.A., Axelrod T.S., Bennett D.P., Chan S., 
Cook K.H., Freeman K.C., Griest K., Marshall S.L., Park H-S., Perlmutter S., Peterson B.A., Pratt M.R., 
Quinn P.J., Rodgers A.W., Stubbs C.W., Sutherland W. Nature, 1993, {\bf 365}, 621 
%
\bibitem[1993]{Aub:a} Aubourg E., Bareyre P., Brehin S., Gros M., Lachieze-Rey M., Laurent B., Lesquoy E., 
Magneville C., Milsztajn A., Moscoso L., Queinnec F., Rich J., Spiro M., Vigroux L., Zylberajch S., 
Ansari R., Cavalier F., Moniez M., Beaulieu J.P., Ferlet R., Grison Ph., Vidal-Madjar 
A., Guibert J., Moreau O., Tajahmady F., Maurice E., Prevot L., Gry C., The Messenger, 
1993a, {\bf 72}, 20
%
\bibitem[1993]{Aub:b} Aubourg E., Bareyre P., Brehin S., Gros M., Lachieze-Rey M.,
Laurent, B., Lesquoy E., Magneville C., Milsztajn A., Moscoso L., Queinnec F., Rich J., Spiro M.,
Vigroux L., Zylberajch S.,
Ansari R., Cavalier F., Moniez M., Beaulieu J.P., Ferlet R.,
Grison Ph., Vidal-Madjar A., Guibert J., Moreau O., Tajahmady F., Maurice E.,
Prevot L., Gry C., Nature, 1993b, {\bf 365}, 623
%
\bibitem[1996]{bb2} Beaulieu, J.P., \etal 1996, in preparation.
\bibitem[1991]{bes} Bessell, M.S. 1991, A\&A, 242, L17
\bibitem[1996]{buc} Buchler, J.R., Kollath, Z., Beaulieu, J.P., \& Goupil, M.J.,
 1996, ApJL, 462, L83.
\bibitem[1985]{cc5} Caldwell, J.A.R. \& Coulson, I.M., 1985, MNRAS, 212, 879
\bibitem[1986]{cc6} Caldwell, J.A.R. \& Coulson, I.M., 1985, MNRAS, 218, 223
\bibitem[1987]{cc7} Caldwell, J.A.R. \& Coulson, I.M., 1987, AJ, 93, 1090
\bibitem[1993]{chi} Chiosi, C., Wood, P.R., \& Capitanio, N., 1993, ApJS, 86, 541.
\bibitem[1995]{dib} Di~Benedetto, G.P., 1995, ApJ, 452, 195.
\bibitem[1991]{fe1} Feast, M.W. 1991, in {\it Observational Tests of
Cosmological Inflation}, eds. Shanks et al., p.147.
\bibitem[1995]{fe2} Feast, M.W. 1995, in {\it Astrophysical Applications of
Stellar Pulsation}, eds. Stobie \& Whitelock, ASP (San Francisco), p.209.
\bibitem[1996]{fer} Ferrarese, L., Freedman, W.L., Hill, R.J., Saha, A., \etal
1996, ApJ, 464, 568
\bibitem[1990]{fma} Freedman, W.L.,\& Madore B.F., 1990, ApJ, 365, 186
\bibitem[1994]{fra} Freedman, W.L., Hughes, S. M., Madore, B.F., \etal 1994a, ApJ, 427, 628
\bibitem[1994]{frb} Freedman, W.L., Madore, B.F., Mould, J. R., \etal 1994b, Nature, 371, 757
\bibitem[1996]{frs} Freedman, W.L., Madore, B.F., \& Kennicutt, R.C. 1996, in 
STScI Symposium on "The Extragalactic Distance Scale", ed. M.Livio, in press.
\bibitem[1993]{gie} Gieren, W.P., Barnes, T.G., \& Moffett, T.J. 1993, ApJ, 418,
135
\bibitem[1986]{gir} Giridhar, S. 1986, J.Astrophys.Astron., 7, 83
\bibitem[1994]{gou} Gould, A. 1994, ApJ, 426, 542
\bibitem[1995]{pet} H\"oflich, P. \& Khokhlov, A. 1996, ApJ, 458, 500
\bibitem[1995]{hss} Huterer, D., Sasselov, D.D., \& Schechter, P.L. 1995, AJ,
  110, 2705
\bibitem[1994]{lst} Laney, C.D., \& Stobie, R.S. 1994, MNRAS, 266, 441.
\bibitem[1991]{maf} Madore, B.F., \& Freedman, W.L. 1991, PASP, 103, 933
\bibitem[1979]{mwf} Martin, W.L., Warren, P.R., \& Feast, M.W. 1979, MNRAS, 188,
139.
\bibitem[1993]{oey} Oey, M.S., \& Kennicutt, R.C. 1993, ApJ, 411, 137
\bibitem[1996]{oli} Olive, K.A., Skillman, E., \& Steigman, G., 1996, preprint
(UMN-TH-1514/96), astro-ph/9611166.
\bibitem[1992]{pag} Pagel, B.E.J., Simonson, E.A., Terlevich, R.J., \&
 Edmunds, M.G. 1992, MNRAS, 255, 325
\bibitem[1994]{pie} Pierce, M.J., Welch, D.L., McClure, R.D., van den Bergh, S.,
Racine, R., \& Stetson, P.B. 1994, Nature, 371, 385
\bibitem[1992]{prs} Press, W.H., Teukolsky, S.A., Vetterling, W.T., \& Flannery,
B.P. 1992, {\it Numerical Recipes}, 2nd Ed., (Cambridge University Press,
Cambridge)
\bibitem[1996]{wpr} Press, W.H. 1996, preprint, astro-ph/9604126.
\bibitem[1995]{ada} Riess, A.G., Press, W.H., \& Kirshner, R.P. 1995, ApJ, 438, L17
\bibitem[1994]{sah} Saha, A., Labhardt, L., Schwengeler, H., Macchetto, F.D.,
 Panagia, N., Sandage, A., \& Tammann, G.A. 1994, ApJ, 425, 14
\bibitem[1994]{san} Sandage, A., Saha, A., Tammann, G.A., Labhardt, L.,
Schwengeler, H., Panagia, N., \& Macchetto, F.D. 1994, ApJ {\bf 423}, L13
\bibitem[1996]{s96} Sandage, A., Saha, A., Tammann, G.A., Labhardt, L., Panagia, N., \&
 Macchetto, F.D. 1996, ApJL, in press.
\bibitem[1980]{sch} Schechter, P.L. 1980, AJ, 85, 801
\bibitem[1982]{s82} Stift, M.J. 1982, A\&A, 112, 149
\bibitem[1995]{s95} Stift, M.J. 1995, A\&A, 301, 776
\bibitem[1988]{sto} Stothers, R. B. 1988, ApJ, 329, 712
\bibitem[1995]{tan} Tanvir, N.R., Shanks, T., Freguson, H.C., \& Robinson,
D.R.T. 1995, Nature, 377, 27
\bibitem[1994]{zar} Zaritsky, D., Kennicutt, R.C., \& Huchra, J. P. 1994, ApJ, 420, 87
\end{thebibliography}
\end{document}